# A Hierarchical Architecture for the Coordination of an Ensemble of Steam Generators

**Stefano Spinelli** [*,**] **Elia Longoni** [*] **Marcello Farina** [*] **Felix Petzke** [***]
**Stefan Streif** [***] **Andrea Ballarino** [**]

[*] *Dipartimento di Elettronica, Informazione e Bioingegneria, Politecnico di Milano, Italy (e-mail: name.surname@polimi.it).*
[**] *Istituto di Sistemi e Tecnologie Industriali Intelligenti per il Manifatturiero Avanzato, Consiglio Nazionale delle Ricerche (e-mail: name.surname@stiima.cnr.it)*
[***] *Department of Electrical Engineering and Information Technology, Technische Universität Chemnitz, Automatic Control and System Dynamics Lab (email: name.surname@etit.tu-chemnitz.de)*

**Abstract:** This work presents a hierarchical architecture for the optimal management of an ensemble of steam generators, which needs to jointly sustain a common load. The coordination of independent subsystems is provided by a multi-layer control scheme. A high-level optimizer computes the optimal shares of production to be allocated to single generators. At medium level, a robust tube-based model predictive control (MPC) is proposed to track the time-varying demand of the ensemble using a centralized, but aggregated model, whose order does not scale with the number of subsystems. At low level, decentralized controllers are in place to stabilize the internal boiler pressure. The control architecture enables the dynamic modification of the ensemble configuration and plug and play operations. Simulation results are reported to demonstrate the potentialities of the proposed approach.

## 1. INTRODUCTION AND PROBLEM STATEMENT

Steam has been widely used as a medium for carrying thermal energy efficiently and plays a central role in production in many sectors, among the others food, textile, chemical, medical, power, heating and transport industries.

In applications with a large steam production demand, where the variation of the request is particularly significant and quick, a flexible and efficient generation is necessary. However, steam generators are mostly inefficient where working close to the lower generation limit and therefore, when operating in scenarios where the demand is extremely fluctuating, the production efficiency may be part-wise very limited. As an alternative to using a single boiler working on a large operating range, it may be preferable to rely on a "virtual" generation system composed by a set of smaller units, operating in parallel to supply what requested, with the possibility to connect/disconnect subsystems quickly and optimally, sharing the effort on available systems in a dynamic way based on the current and/or forecasted demand. The main objective of this work is the proposal of a hierarchical control scheme for the optimal management of a group of steam generators that work in a parallel configuration to sustain a cumulative steam demand.

Different industrial applications are characterized by the need of coordination of independent/interdependent subsystems towards a main goal, e.g. electrical generation systems, microgrids, building heating and cooling systems or water distribution networks. All these examples share some features, which make the corresponding management problem challenging from the methodological standpoint. They are complex plants where several similar systems operate in parallel to jointly produce a common product. These systems operate in a limited range and are forced to cooperate in a scenario of limited shared resources to supply an overall demand.

In the recent years, some efforts have been devoted to provide solutions to this problem. Decentralized, distributed, and hierarchical methods have many advantages over centralized ones, in view of their flexibility, robustness (e.g., to system changes and demand variations), and scalability. In this work we focus our attention on hierarchical methods, that are considered as the elective choice for optimal supervision and coordination of the system ensembles.

With particular reference to systems where limitations and constraints play a key role, different algorithms have been proposed recently, e.g., based on model predictive control (MPC) (Scattolini, 2009; Barcelli et al., 2010; Picasso et al., 2016; Farina et al., 2018; Petzke et al., 2018) and reference governors (RG) (Garone et al., 2017; Kalabić et al., 2012).

In this work, the proposed hierarchical architecture is inspired by (Petzke et al., 2018) and (Farina et al., 2018). More specifically, we consider a group of $N_g$ steam generators that work in a parallel configuration to sustain a cumulative steam demand, $\bar{q}_s^{\mathrm{Dem}}$. We aim to fulfill the required steam flow rate, with the minimum amount of fuel gas and optimizing the contribution of each boiler to the overall demand.

We assume that the steam generators - although different in dimensions, steam and firing rate as well as efficiency - are homogeneous dynamical systems, i.e. sets of *similar* subsystems - in terms of input and output. Each subsystem $i$ is a water-tube boiler, formed by a tube coil where a pressurized water, denoted as feed-water $q_{f,i}$, is forced to flow by a displacement pump and heated by a natural gas burner, whose flow rate is $q_{g,i}$. The heat, transmitted to the flowing fluid, induces a phase transition of the feed-water into steam. The steam flow rate generated by boiler $i$ is $q_{s,i}$. The ensemble and the single subsystems are subject to input and output constraints. We assume that convex and compact sets $\mathscr{U}_i$, $\mathscr{Y}_i$, $\bar{\mathscr{U}}$ and $\mathscr{Y}$ are defined in such a way that

$$q_{s,i} \in \mathscr{U}_i \qquad q_{g,i} \in \mathscr{Y}_i \tag{1a}$$

$$\bar{q}_s = \sum_{i=1}^{N_g} q_{s,i} \in \bar{\mathscr{U}} \qquad \bar{q}_g = \sum_{i=1}^{N_g} q_{g,i} \in \mathscr{Y} \tag{1b}$$

The proposed scheme consists of three layers.

The *top layer*, inspired by the one discussed in (Farina et al., 2018), computes the optimal shares of production to be allocated to each boiler based on the request profile $\bar{q}_s^{\text{Dem}}$ minimizing the operation cost, i.e. the fuel usage. This includes the possibility to activate and deactivate boilers in the ensemble. In particular, defining with $\bar{q}_s$ the total produced steam, sharing factor $\alpha_i$ are defined such that

$$q_{s,i} = \alpha_i \bar{q}_s \quad \text{with } \alpha_i \in [0,1] \quad \text{and } \sum_{i=1}^{N_g} \alpha_i = 1 \tag{2}$$

This layer has remarkable differences with respect to (Farina et al., 2018). For instance, in this work a procedure is proposed to limit the conservativeness of the overall scheme, but at the same time to avoid inconsistencies and constraint violation at the lower levels.

At the *medium control layer*, we adopt the scheme proposed in (Petzke et al., 2018), that allows to track the overall demand $\bar{q}_s$ using an MPC algorithm applied to the aggregate low-order model of the system ensemble in a scalable way. At the same time, this allows to determine the local steam requests $\bar{q}_{s,i} = \alpha_i \bar{q}_s$ for each subsystem.

At the *lowest layer*, we use a decentralized set of proportional-integral (PI) controllers, i.e., the ones currently used in industrial practice, to track the individual requests and regulating the internal pressure. The lower layer can be possibly enhanced in line with (Petzke et al., 2018) with local shrinking horizon MPC controllers, operating at faster sampling time, to improve the performance of the control of the ensemble. The models used for low-level control are nonlinear ones, derived from physical equations. However, the corresponding (closed loop) models used at the higher hierarchical layers are affine and identified from data extracted by the nonlinear simulators.

## 2. BOILER MODELS AND LOW-LEVEL CONTROLLER

In this section we present the dynamical model of the steam generators and the PI controller applied at the low hierarchical level. The model and the controller structure are the same for each boiler $i$ of the ensemble; for notational simplicity, the index $i$ will be dropped when clear from the context.

### 2.1 The nonlinear boiler model

The continuous-time nonlinear dynamical model is derived from the drum-boiler model presented in Åström and Bell (2000), whose equations are adapted to the configuration of the steam generator: it differs from the drum-boiler as the drum is absent and no accumulation exists in the water tubes. In particular, as the feed-water flow-rate $q_f$ is forced through the heated tubes, this is either totally or partially transformed into steam. Therefore, we can write:

$$q_f = q_s + q_w \tag{3}$$

where $q_s$ is the steam flow-rate. The residual part $q_w$ remains in liquid phase and is assumed to be at saturated temperature.

The $i$-th steam generator is characterized by a nonlinear dynamic model $\mathscr{S}_i^{\text{NL, OL}}$:

$$\dot{p} = \frac{1}{\phi}\left(\eta \lambda_{LHV} q_g + q_f(h_f - h_w) - q_s(h_s - h_w)\right) \tag{4}$$

$$\dot{V}_w = \frac{1}{(\rho_w - \rho_s)}\left(\frac{\partial \rho_w}{\partial p} V_w + \frac{\partial \rho_s}{\partial p} V_s\right)\dot{p} \tag{5}$$

where

$$\phi = V_s\left(h_s \frac{\partial \rho_s}{\partial p} + \rho_s \frac{\partial h_s}{\partial p}\right) + V_w\left(h_w \frac{\partial \rho_w}{\partial p} + \rho_w \frac{\partial h_w}{\partial p}\right) + V_T + m_T c_p \frac{\partial T_s}{\partial p} - \left(\frac{\partial \rho_w}{\partial p} V_w + \frac{\partial \rho_s}{\partial p} V_s\right)\frac{(\rho_w h_w - \rho_s h_s)}{(\rho_w - \rho_s)} \tag{6}$$

In equations (4)-(6), $\rho, h,$ and $T$ are respectively density, enthalpy and temperature that are function of internal pressure $p$. The subscripts $_{f,g,s,w}$ refer to feed-water, fuel gas, steam, internal water. Subsystem specific parameters are: the burner efficiency, $\eta$, the gas low heat value $\lambda_{LHV}$, the total tubes internal volume, $V_T$, mass, $m_T$ and specific heat coefficient, $c_p$.

In the nonlinear dynamical model (4)-(5), the states $p$ and $V_w$ are the internal pressure $p$ and the water volume $V_w$. The inputs are $q_f$ and $q_g$ while the steam demand $q_s$ is considered, at the low-level, as a disturbance term.

### 2.2 Low-level controller

We include an embedded controller in each subsystem $\mathscr{S}_i^{\text{NL, OL}}$ that operates on $q_f$ and $q_g$ (or, more specifically, on the local input variables $q_{f,i}$ and $q_{g,i}$) to maintain the pressure at the set-point level and the water volume $V_w$ constant. Note that, however, commercially-available boilers are currently already provided with low-level controllers. For this reason, in this paper we apply the realistic assumption that the latter are designed based on the industrial standard configuration. More specifically, we assume that a PI regulator **R** and a disturbance compensator **C** act on the fuel flow rate and on the feed-water flow rate, respectively, as depicted in Figure 1. **R** is a feedback controller regulating the pressure to a set-point $p_{sp}$, while the compensator **C** forces, with an open-loop action, the feed-water to follow the steam demand. As a result, the $i$-th boiler, controller at low-level, can be described as a nonlinear dynamic model $\mathscr{S}_i^{\text{NL, CL}}$, in short denoted as

$$q_{g,i} = \mathscr{S}_i^{\text{NL, CL}}(q_{s,i}) \tag{7}$$

The steam flow rate can be accounted for as input of the control system while the gas flow rate will be considered as an output,

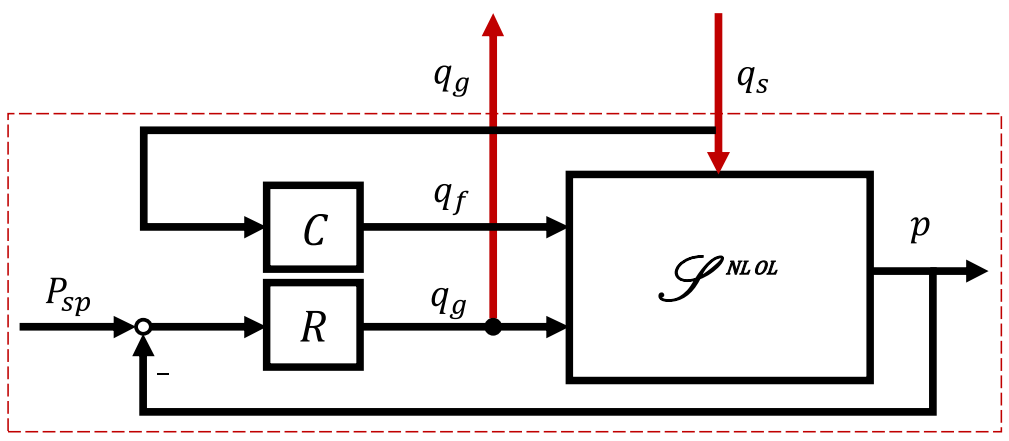

Fig. 1. Closed-loop Steam Generator block diagram. The stabilized system is framed with the red dashed line, highlighting the new input-output pair considered by the system (7)

as shown in Figure 1. This reverse vision of the boiler, with respect to the actual physical flows of fuel and steam, permits to formalize the problem in the framework of hierarchical control of ensemble systems.
Since the pressure regulator keeps the system close to the nominal working plant conditions, the dynamics of $\mathscr{S}_i^{\text{NL, CL}}$ are very close to the dynamics of a linear system. This is witnessed, for example, by Figure 2, where the input/output static map of the system is reported: the dataset 1 of historical static data (gray dots) is compared with the dataset 2 generated simulating the response of the system $\mathscr{S}_i^{\text{NL, CL}}$ with a multiple step input profile (red stars). The linear regression model is shown as dashed line. In the next section we will identify the corresponding affine model. This will be particularly useful for the design of the higher hierarchical control layers.

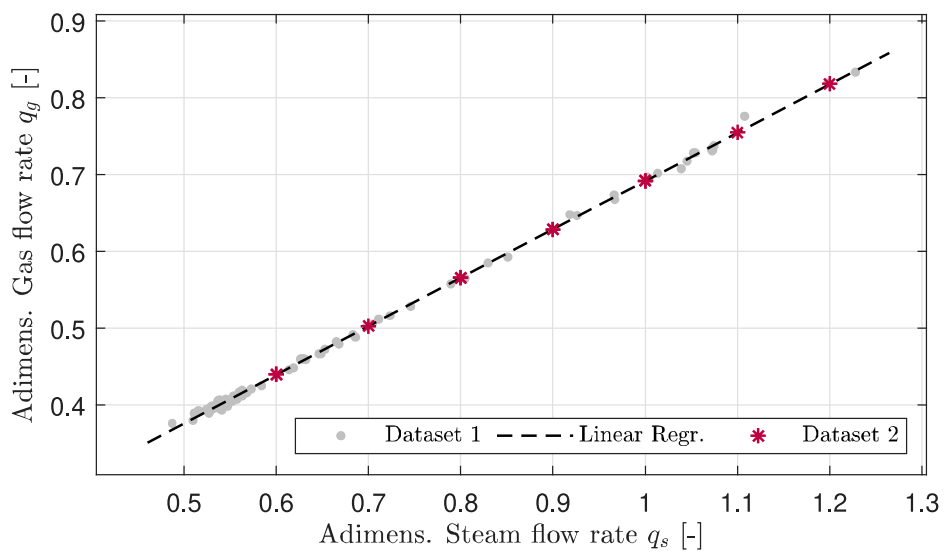

Fig. 2. Input-output static map of the controlled generator

### 2.3 Linear low-level closed loop system

In this section we identify an affine model $\mathscr{S}_i^{\text{L, CL}}$ using the controlled nonlinear system $\mathscr{S}_i^{\text{NL, CL}}$ to generate the data. The model is identified with a "fast" sampling time $\tau$ and $\kappa$ is the corresponding time index. We define $y(\kappa) = q_{g,i}(\kappa\tau)$ and we assume that the input is defined in a sample-and-hold way, i.e., $u(\kappa) = q_{s,i}(t)$ is constant for all $t \in [\kappa\tau, (\kappa+1)\tau)$. Model $\mathscr{S}_i^{\text{L, CL}}$ is first identified as a discrete-time transfer function (plus constant) of the type

$$y(\kappa) = \frac{\sum_{j=1}^{n_\text{b}}(b_j z^{-j})}{1+\sum_{j=1}^{n_\text{f}}(f_j z^{-j})} u(\kappa) + \gamma \tag{8}$$

where $\gamma$ is the identified bias when $u(\kappa) = 0$ in steady-state.
The corresponding state-space form

$$\mathscr{S}^{\text{L, CL}} : \begin{cases} x(\kappa+1) = Ax(\kappa) + Bu(\kappa) \\ y(\kappa) = Cx(\kappa) + \gamma \end{cases} \tag{9}$$

is obtained by defining $x(\kappa) = [y(\kappa), y(\kappa-1), ..., y(\kappa - n_\text{f} + 1), u(\kappa-1), ..., u(\kappa - n_\text{b} + 1)]^T \in \mathbb{R}^n$, where $n = n_\text{f} + n_\text{b} - 1$, and setting $B = \begin{bmatrix} b_1 & 0_{1\times(n_\text{f}-1)} & 1 & 0_{1\times(n_\text{b}-2)} \end{bmatrix}^T$, $C = [1 \ 0 \ \dots \ 0]$, and

$$A = \left[\begin{array}{cc|cc} -f_1 \ \dots \ -f_{n_\text{f}-2} & -f_{n_\text{f}} & b_2 \ \dots \ b_{n_\text{b}-1} & b_{n_\text{b}} \\ I_{n_\text{f}-1} & 0_{(n_\text{f}-1)\times 1} & \multicolumn{2}{c}{0_{(n_\text{f}-1)\times(n_\text{b}-1)}} \\ \hline \multicolumn{2}{c|}{0_{(n_\text{b}-1)\times n_\text{f}}} & 0_{1\times n_\text{b}-2} & 0 \\ & & I_{n_\text{b}-2} & 0_{(n_\text{b}-2)\times 1} \end{array}\right]$$

## 3. MEDIUM-LEVEL AND HIGH-LEVEL CONTROLLERS

At the higher hierarchical levels, the subsystems (controlled at low level) are assumed to be described by the state-space models (9), which are possibly different from each other. For instance, for subsystem $i = 1, \dots, N_\text{g}$

$$\mathscr{S}_i^{\text{L, CL}} : \begin{cases} x_i(\kappa+1) = A_i x_i(\kappa) + B_i u_i(\kappa) \\ y_i(\kappa) = C_i x_i(\kappa) + \gamma_i \end{cases} \tag{10}$$

As in Farina et al. (2018), it is assumed that the following properties hold.

*Assumption 1.* It is required that: (i) $A_i$ is Schur stable; (ii) the system is squared, i.e., $m = p$; (iii) $\det(C_i(I_n - A_i)^{-1}B_i) \neq 0$.

All the identified systems $\mathscr{S}_i^{\text{L, CL}}$ respect the assumptions above: in particular, in the present work, we have that $m = p = 1$. As shown in Figure 3 and as also discussed in Section 1, the control architecture proposed in the present work allows to compute the individual inputs (i.e., local steam flowrate) as $u_i = \alpha_i \bar{u}$ where

- the parameters $\alpha_i$, $i = 1, \dots, N_\text{g}$ are the sharing factors of the boilers and are computed by the high-level static optimization layer;
- $\bar{u}$ is the input to the aggregated ensemble model and is computed by a dynamic optimal reference tracking problem at medium level, operating on a slow time scale.

### 3.1 High-level Optimization

The high-level (HL) is devoted to the optimization of the sharing factors $\alpha_i$, which define the partition of the overall demand among the subsystems of the ensemble. This optimization layer considers the functioning range of each subsystem in the ensemble to ensure the best share of resources and the minimization of the associated operating cost. In this work a static optimization layer is discussed, although this framework can be extended to include dynamics and day-ahead predictions. In a static environment, we define $g_i = C_i(I_{n_i} - A_i)^{-1}B_i$.
We assume that $\bar{q}_\text{s}^\text{Dem}$ is given. Our scope is to compute, at the same time, the corresponding shares $\alpha_i$ and a feasible steady-state overall steam production value $\bar{u}_\text{ss}$, as close as possible to the demand $\bar{q}_\text{s}^\text{Dem}$, but which, at the same time, allows to fulfill constraints (1), where $\mathscr{U}_i = [u_{min,i}, u_{max,i}]$, $\mathscr{Y}_i = [y_{min,i}, y_{max,i}]$, $\bar{\mathscr{U}} = [\bar{u}_{min}, \bar{u}_{max}]$, and $\bar{\mathscr{Y}} = [\bar{y}_{min}, \bar{y}_{max}]$. We include a further constraint to enforce, possibly in a conservative way, consistency with the medium level controller. More specifically, at medium level, a limitation on the variation of the input value between two consecutive steps is enforced to be lower (in absolute value) than $\Delta\bar{u}$, as it will be better discussed later in the paper, for all active boilers (in this respect, we introduce the integer variable $\delta_i$, being $\delta_i = 1$ if the boiler $i$ is active, while $\delta_i = 0$ otherwise). More specifically, we require, for each subsystem, that $|\alpha_i\bar{u}_{\rm ss} - \delta_i\alpha_i^{\rm old}\bar{u}_{\rm ss}^{\rm old}| \le \alpha_i^{\rm old}\Delta\bar{u}$, where $\alpha_i^{\rm old}$ and $\bar{u}_{\rm ss}^{\rm old}$ are the values of $\alpha_i$ and $\bar{u}_{\rm ss}$ applied before the optimization is solved.
A mixed-integer program with bilinear inequality constraints is formulated, also allowing for the possibility of having different weights $\lambda_i$ for each subsystems, e.g., related to the cost of using each single boiler. The optimization problem reads

$$\min_{\alpha_i,\delta_i,\bar{u}_{\rm ss}} \sum_{i=1}^{N_{\rm g}} \lambda_i\left(g_i\alpha_i\bar{u}_{\rm ss} + \delta_i\gamma_i\right) + \bar{\lambda}\left(\bar{u}_{\rm ss} - \bar{q}_{\rm s}^{\rm Dem}\right)^2 \tag{11}$$

$$\begin{aligned}
\text{s.t.}\quad & \sum_i \alpha_i = 1\\
& \bar{u}_{min} \le \bar{u}_{ss} \le \bar{u}_{max}\\
& \bar{y}_{min} \le \sum_{i=1}^{N_{\rm g}}\left(g_i\alpha_i\bar{u}_{\rm ss} + \delta_i\gamma_i\right) \le \bar{y}_{max}\\
\text{and, for all } & i = 1,\ldots,N_{\rm g}\\
& u_{min,i}\delta_i \le \alpha_i\bar{u}_{ss} \le u_{max,i}\delta_i\\
& y_{min,i}\delta_i \le g_i\alpha_i\bar{u}_{ss} + \delta_i\gamma_i \le y_{max,i}\delta_i\\
& -\alpha_i^{\rm old}\Delta\bar{u} \le \alpha_i\bar{u}_{\rm ss} - \delta_i\alpha_i^{\rm old}\bar{u}_{\rm ss}^{\rm old} \le \alpha_i^{\rm old}\Delta\bar{u}\\
& 0 \le \alpha_i \le 1 \quad\text{and}\quad \delta_i \in \{0,1\}
\end{aligned}$$

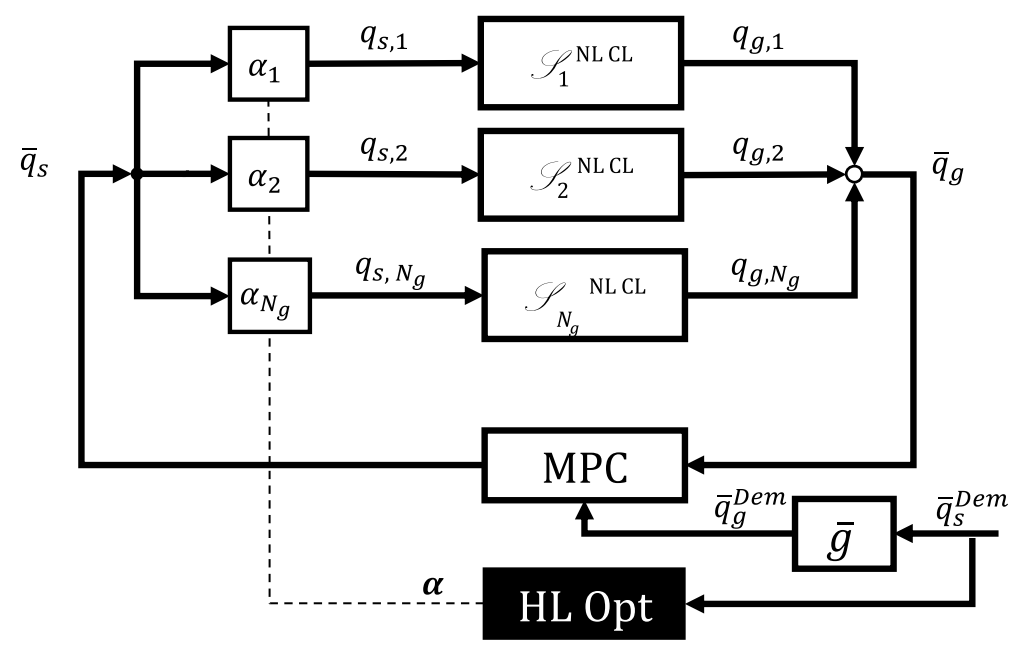

Fig. 3. Steam Generator ensemble and hierarchical scheme.

The high-level optimization can be either event-based or cyclically executed. In event-based mode, the triggering can be done, e.g., when the steam demand varies significantly with respect to the value used in the previous run of the high-level itself, i.e., when $|\bar{q}_{\rm s}^{\rm Dem} - \bar{q}_{\rm s}^{\rm Dem,old}| \ge \Delta\bar{q}_{\rm s}^{\rm Dem,threshold}$, for a given case-dependent value of $\Delta\bar{q}_{\rm s}^{\rm Dem,threshold}$.

### 3.2 Medium-level control

*Ensemble model* For control design at medium level, we first need to devise a model of the boiler ensemble. To do so, we adopt the procedure sketched in (Petzke et al., 2018), and we first define the *reference* dynamics for each subsystem. In practice, we need: (i) to define a possibly reduced state of the $i$-th reference model as $\hat{x}_i = \beta_i x_i$ where $\beta_i \in \mathbb{R}^{\hat{n}\times n_i}$ is a suitable map, where $\hat{n} \le n_i$; (ii) to describe the evolution of the state variable $\hat{x}_i$ using the alternative model

$$\hat{\mathscr{S}}_i : \begin{cases} \hat{x}_i(\kappa+1) = \hat{A}\hat{x}_i(\kappa) + \hat{B}_i u_i(\kappa) + \hat{w}_i(\kappa)\\ y_i(\kappa) = \hat{C}\hat{x}_i(\kappa) + \hat{\gamma}_i \end{cases} \tag{12}$$

where the term $\hat{w}_i(\kappa)$ embeds the error due to the mismatch between model (12) and (10). In (Petzke et al., 2018) it is shown that, if the rate of change of $\bar{u}$ between consecutive steps is bounded, then it is possible to guarantee that $\hat{w}_i(\kappa)$ is bounded. It is important to point out that the state and output matrices $\hat{A}$ and $\hat{C}$, respectively, as well as the model order $\hat{n}$, are the same for all subsystems' reference models. On the other hand, matrices $\hat{B}_i$ must be properly selected. The most convenient choice consists of selecting $\hat{A}$, $\hat{B}_i$ and $\hat{C}$ with a similar canonical structure of $A_i$, $B_i$ and $C_i$.
In this case it is sufficient to define $\beta_i \in \mathbb{R}^{\hat{n}\times n_i}$ as a suitable selection matrix, i.e., whose rows are basis vectors of the canonical space and where $\hat{n} \le n_i$. The reference model (12) must satisfy the so-called *gain consistency* conditions with respect to (10), that in this case are verified by simply setting

$$\hat{\gamma}_i = \gamma_i \tag{13}$$

$$\hat{b}_{i,1} = \frac{\sum_{j=1}^{n_{\rm b}} b_{i,j}}{1+\sum_{j=1}^{n_{\rm f}} f_{i,j}}\left(1+\sum_{j=1}^{\hat{n}_{\rm f}}\hat{f}_j\right) - \sum_{j=2}^{\hat{n}_{\rm b}}\hat{b}_j \tag{14}$$

where $b_{i,j}$, $j = 1,\ldots,n_{\rm b}$ and $f_{i,j}$, $j = 1,\ldots,n_{\rm f}$ are the parameters that characterize the $i$-th model (12).

The state of the ensemble dynamical model $\bar{\mathscr{S}}$ is defined considering the states of the single active boilers, i.e., the ones where $\delta_i = 1$. This is done under the assumption that, when a boiler is switched off (i.e., when $\delta_i$ is set to 0), its steam production is inactivated and the steam produced during the switch off transient is diverted from the ensemble output. Accordingly, $\bar{x} = \sum_i^{N_g}\delta_i\hat{x}_i$, the input is $\bar{u}$, and the output is $\bar{y} = \sum_i^{N_g}\delta_i y_i$. Considering the reference models (12), we can write

$$\bar{\mathscr{S}} : \begin{cases} \bar{x}(\kappa+1) = \hat{A}\bar{x}(\kappa) + \bar{B}\bar{u}(\kappa) + \bar{w}(\kappa)\\ \bar{y}(\kappa) = \hat{C}\bar{x}(\kappa) + \bar{\gamma} \end{cases} \tag{15}$$

where $\bar{B} = \sum_i^{N_g}\alpha_i\hat{B}_i$, $\bar{\gamma} = \sum_i^{N_g}\delta_i\hat{\gamma}_i$, and $\bar{w} = \sum_i^{N_g}\delta_i w_i$. We also define the static gain of the ensemble as $\bar{g} = \sum_i^{N_g} g_i\alpha_i$.
At medium level, the reference tracking controller may operate with a slow sampling time $T = \nu\tau$, where $\nu \in \mathbb{N}$, whose corresponding time index is $k$. This requires to define the ensemble variables at the new timescale as follows. The input is $\bar{u}^{[T]}(k)$, defined such that $\bar{u}(\kappa) = \bar{u}^{[T]}(k)$ for all $\kappa = k\nu,\ldots,(k+1)\nu-1$); also, we define $\bar{x}^{[T]}(k) = \bar{x}(k\nu)$ and $\bar{y}^{[T]}(k) = \bar{y}(k\nu)$. Accordingly, the system (16) is re-sampled, and its slow timescale dynamics is:

$$\bar{\mathscr{S}}^{[T]} : \begin{cases} \bar{x}^{[T]}(k+1) = \hat{A}^{[T]}\bar{x}^{[T]}(k) + \bar{B}^{[T]}\bar{u}^{[T]}(k) + \bar{w}^{[T]}(k)\\ \bar{y}^{[T]}(k) = \hat{C}^{[T]}\bar{x}^{[T]}(k) + \bar{\gamma} \end{cases} \tag{16}$$

where $\hat{A}^{[T]} = \hat{A}^{\nu}$, $\bar{B}^{[T]} = \sum_{j=0}^{\nu-1}\hat{A}^j\bar{B}$, and $\bar{w}^{[T]}(k)$ is defined consistently.
We define the set where $\bar{w}(k)$ lies as $\bar{\mathscr{W}}$. Its computation can be done following the procedure discussed in (Petzke et al., 2018) from set $\Delta\bar{\mathscr{U}} = [-\Delta\bar{u}, \Delta\bar{u}]$, where $\Delta\bar{u}$ is defined previously in Section 3.1.

*Medium-level controller design* The objective of the medium level MPC is to track the global fuel flow rate target $r = \bar{q}_{\rm g}^{Dem}$, based on the ensemble configuration, defined by the sharing factors optimized at high level.
For all the time steps $k$, the medium level is also committed to enforce the constraints (1), i.e.,

$$\bar{u}^{[T]}(k) \in \bar{\mathscr{U}} \tag{17a}$$

$$\bar{y}^{[T]}(k) \in \bar{\mathscr{Y}} \tag{17b}$$

and, for all $i = 1,\ldots,N_{\rm g}$

$$u_i(k\nu) = \alpha_i\bar{u}^{[T]}(k) \in \mathscr{U}_i \tag{17c}$$

$$y_i(k\nu) \in \mathscr{Y}_i \tag{17d}$$

Also, for consistency, we need to ensure that for all $i = 1,\ldots,N_{\rm g}$

$$\alpha_i(k)\bar{u}^{[T]}(k) - \alpha_i(k-1)\bar{u}^{[T]}(k-1) \in \alpha_i(k-1)\Delta\bar{\mathscr{U}} \tag{17e}$$

To manage this, the MPC is a robust offset-free tracking algorithm based on the formulation presented in Betti et al. (2013). To design the offset-free state-feedback MPC, the ensemble model is augmented and rewritten in the velocity form. This robust MPC algorithm in velocity form, among other things, has the advantage to easily enforce the constraints (17e). Furthermore, thanks to the tube-based MPC approach, by tightening opportunely the constraints, the MPC problem guarantees the feasibility of the actual controller while considering the unperturbed system in the computation of the control action. Finally, as also done in Limon et al. (2008), in the algorithm proposed in Betti et al. (2013), the corresponding optimization program to be solved at each time step $T$ is enhanced with the additional optimization variable $\hat{r}$, which is defined as the closest feasible set point to $r$, to guarantee feasibility when set-points change. A final remark is due. While constraints (17a)-(17c) and (17e) can be directly written as constraints on the ensemble state variables and inputs (and therefore do not involve any implementation problem), the constraints (17d) are related to local outputs. Note that, in (Petzke et al., 2018) they are not considered. However, in our application scenario, they play a key role, since they represent limitations in the gas available to each burner. Therefore, here, they must be considered. To enforce them, two possible choices are possible: (i) to consider explicitly models (10) to generate constraints on variables $y_i$; (ii) to replace (17d) with the simplified "quasi steady-state" version

$$g_i \alpha_i \bar{u}^{[T]}(k) + \gamma_i \in \bar{\mathscr{Y}}_i$$

where $\bar{\mathscr{Y}}_i$ is computed by suitably tightening set $\mathscr{Y}_i$.

## 4. SIMULATIONS

We consider a use case with $N_g = 5$ steam generators - operating at nominal pressure of 57 bar - serving a common load. These boilers differ slightly among each other, in the dimensions, operation range - minimum/maximum generated steam - and burner efficiency: respective parameters are listed in Table 1.

Table 1. Boiler parameters

| **Boiler n** | 1 | 2 | 3 | 4 | 5 |
|---|---|---|---|---|---|
| $V_T$ $[m^3]$ | 1.21 | 1.15 | 1.28 | 1.14 | 1.32 |
| $m_T$ $(\times 10^3)[kg]$ | 5.499 | 5.220 | 5.830 | 5.060 | 5.995 |
| $\eta$ $[-]$ | 0.90 | 0.92 | 0.89 | 0.95 | 0.99 |
| $q_s$ Min $[kg/s]$ | 0.1 | 0.092 | 0.089 | 0.095 | 0.099 |
| $q_s$ Max $[kg/s]$ | 1.264 | 1.16 | 1.125 | 1.20 | 1.25 |
| $q_g$ Min $[kg/s]$ | 0.1251 | 0.1273 | 0.1295 | 0.1253 | 0.1227 |
| $q_g$ Max $[kg/s]$ | 0.8588 | 0.8435 | 0.8458 | 0.8414 | 0.8389 |
| $\lambda$ $[-]$ | 100 | 130 | 120 | 70 | 80 |

In addition, the system is characterized by the following global constraints $\bar{\mathscr{Y}} = [0.1227, 4.220]$ $[kg/s]$ and $\bar{\mathscr{U}} = [0.089, 6.0]$ $[kg/s]$, determined by constraints of the distribution network.
The low-level controllers depicted **C** and **R** in Figure 1 are defined as described in Section 2.2: the compensator and regulator operate at discrete time with a sampling time, $\tau = 10$ s and their parameters are tuned to stabilize the system with a settling time of 120 s. We assume that all systems have the same low level controllers, whose parameters are tuned by standard procedures and take the following values: **R** with $\mathrm{K_P} = 0.87$ and $\mathrm{K_I} = 3.54\,10^{-4}$, while **C** with $\mathrm{K_P} = 0.31$ and $\mathrm{K_I} = 0.1$.
The close-loop nonlinear model is used to generate the dataset for the identification of the discrete-time linear polynomial model (8), with $\tau = 10$ s, $\mathrm{n_f} = 3$, $\mathrm{n_b} = 2$ and $\mathrm{n_k} = 1$. The same setting is used for each boiler, so that systems $\mathscr{S}_i^{L,CL}$ have the same order $n$. The comparison of the dynamic response of the nonlinear model $\mathscr{S}_i^{\mathrm{NL,\ CL}}$ and the identified linear one $\mathscr{S}_i^{L,CL}$ is presented in Figure 4 for boiler $i = 1$. The matrices of the 1st steam generator are chosen for defining the reference model. In Figure 5, the comparison of the step response of each system $\mathscr{S}_i^{L,CL}$ with its reference model $\hat{\mathscr{S}}_i$ is shown: the gain consistency conditions (13)-(14) guarantee that at steady state the actual and reference models reach the same value.
Thanks to the reference models $\hat{\mathscr{S}}_i$, the model $\bar{\mathscr{S}}$ of the ensemble is derived using (16) and re-sampled at the medium-layer time scale, with $T = 30$ s.
The robust MPC is designed considering that the disturbance $\bar{w}$ is such that $\|\bar{w}\|_\infty \leq 4 \times 10^{-2} [kg/s]$: the set $\bar{\mathscr{W}}$ is evaluated by imposing the maximum variation of the input equal to $\Delta\bar{u} = 0.5[kg/s]$.
A simulation shows the results of the proposed control architecture when a piece-wise constant demand is given. It is worth noting that, as reported in Figure 3, the reference trajectory is naturally given in terms of steam demand $\bar{q}_s^{\mathrm{Dem}}$ and converted into equivalent gas target using the static gain of the ensemble $\bar{q}_g^{\mathrm{Dem}} = \bar{g} \cdot \bar{q}_s^{\mathrm{Dem}}$. As shown in Figure 6, the reference trajectory for the the ensemble is further modulated at high and medium levels: its variation is restrained at high level by considering the implication of the feasibility of the medium predictive controller, as discussed in Section 3.1.

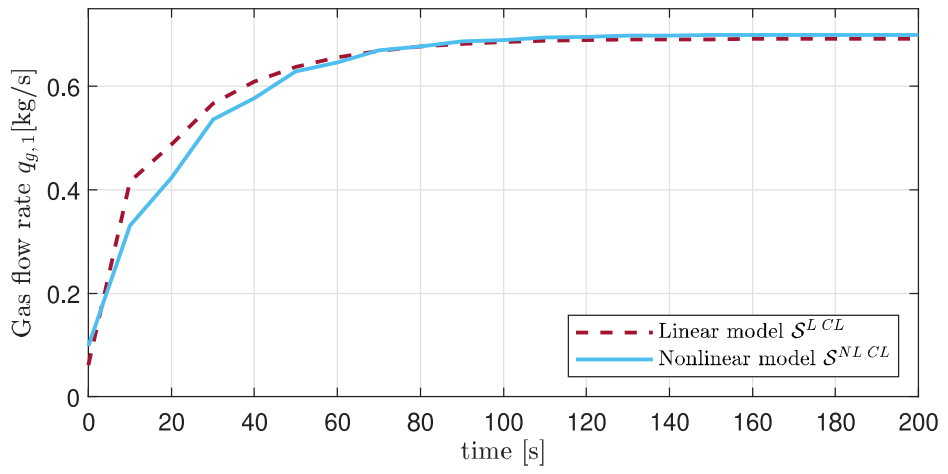

Fig. 4. Comparison of step-reponse of nonlinear model $\mathscr{S}_i^{\mathrm{NL,\ CL}}$ (solid) and linearized model $\mathscr{S}_i^{L,CL}$ (dashed) for Boiler 1.

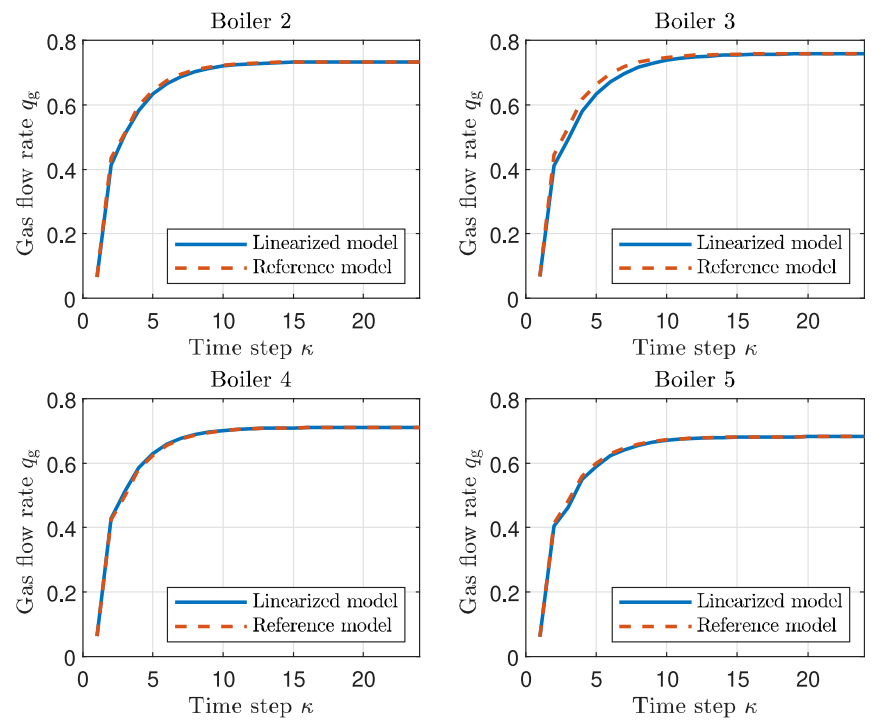

Fig. 5. Comparison of the step-response of actual (solid) and reference models (dashed). Comparison for Boiler 1 not reported since it is chosen as reference model.

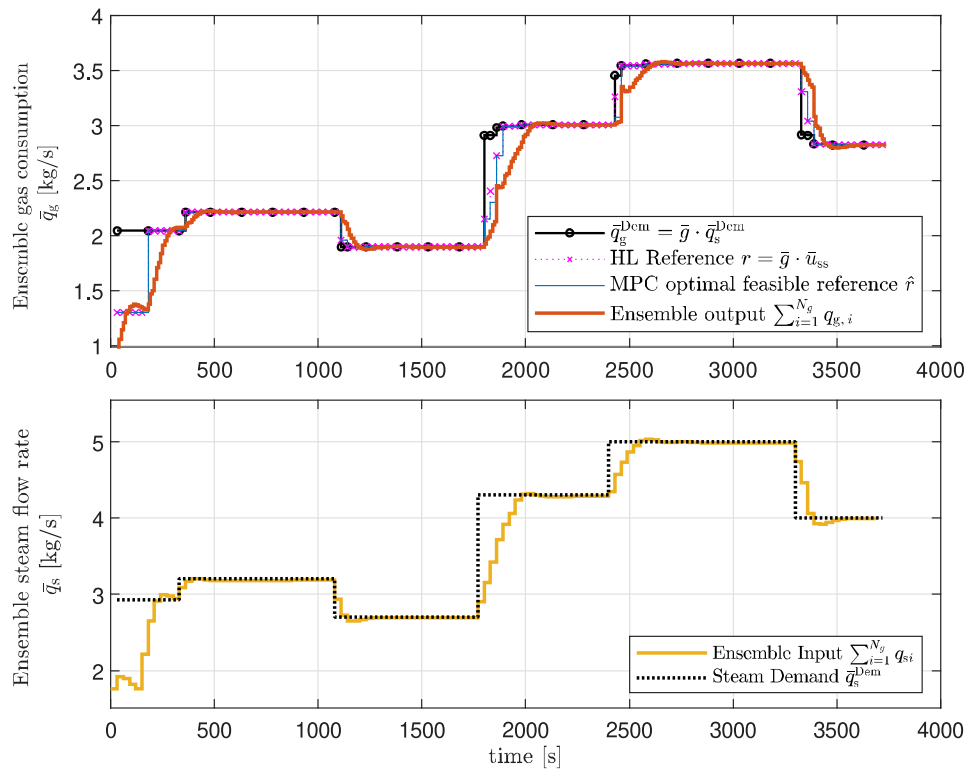

Fig. 6. *Top Graph* - The ensemble gas consumption (thick red solid line) is shown tracking a time-varying load. The overall gas demand, computed with the static relation (black line with circles), is shown with the feasible steady state steam production (magenta dotted line). The medium-MPC reference is shown as thin blue solid line. *Bottom Graph* - The ensemble steam generation (ocher solid line) is shown with the time-varying steam demand, (black dotted line).

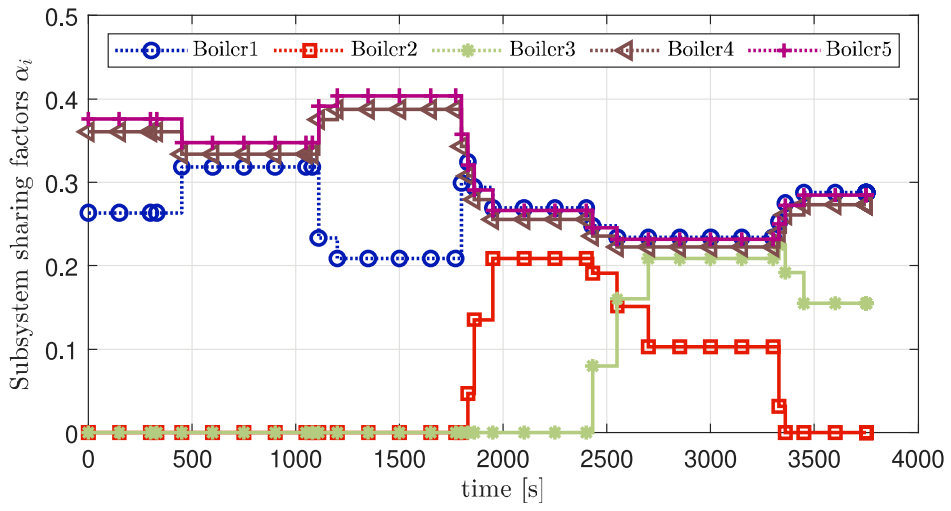

Fig. 7. Sharing factors $\alpha_i$ computed by top level optimization.

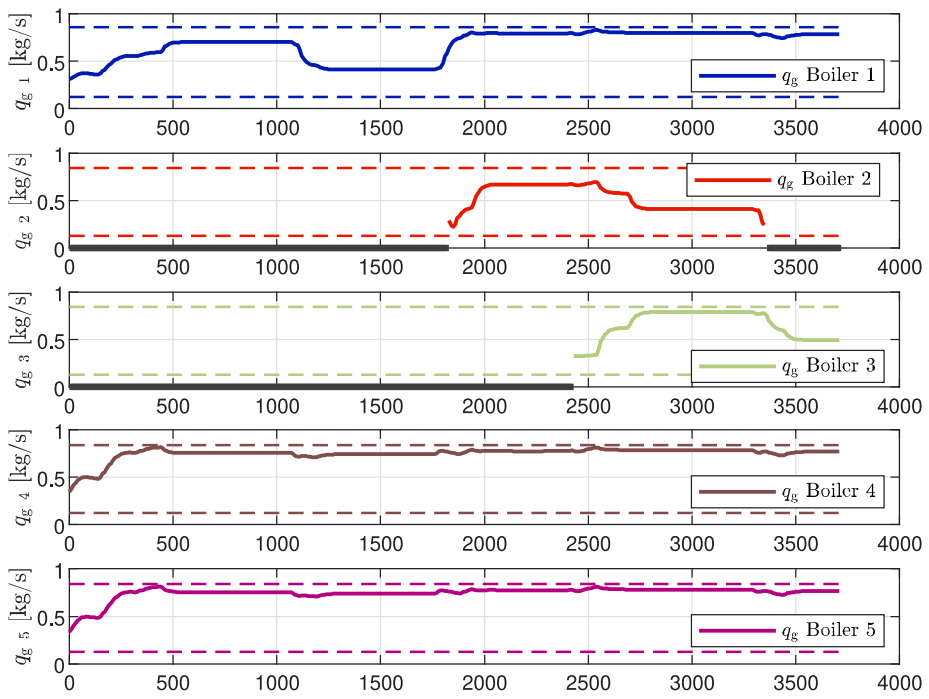

Fig. 8. In each subplot, the gas flow rate $q_{g,i}$ of the single subsystem is shown. Unplugged boilers are shown with a bold gray line below the minimum level.

The high-level optimization is executed with a slow cycle time, a multiple of the medium level controller, i.e. $T_{HL} = 5T$. However, the optimization is also triggered when the distance from the final target is above a certain threshold, $\Delta\bar{q}_s^{\mathrm{Dem,threshold}} = 3 \times 10^{-2}$. The high level concurrently decides the shares, the activation/deactivation of generators and the closest reachable steady-state steam value $\bar{u}_{ss}$ used to compute the reference for the lower MPC level, $r = \bar{g} \cdot \bar{u}_{ss}$, to safeguard its feasibility.

As shown in Figure 7, when the global steam demand rises, the generators are added to the ensemble based on the subsystem efficiency rank, but also to the associated operating cost $\lambda_i$. When the demand slightly changes as in the first half of the simulation, sharing factors are just adapted to improve the ensemble operating efficiency. Instead, a larger increase in the demand, as at $t = 1800$ s, induces a variation of the ensemble configuration, shown by the activation of the boiler 2. To respond to a further increase of the demand, at $t = 2450$ s, the boiler 3 is plugged into the ensemble, then when the demand drops again, boiler 2 is unplugged. When the share factors change or the ensemble configuration is modified by the introduction or removal of a subsystem, the ensemble model is recomputed following (16) and the MPC state $\bar{x}^{[T]}$ and the previous optimal input $\bar{u}_{k-1}^{*[T]}$ are modified in order to be consistent with the successive configuration of the ensemble. The optimal steam demand for the ensemble $\bar{q}_s$ is computed by the medium level MPC to track the reference trajectory. Based on the shares given by the high layer, the subsystem steam flow rate $q_{s,i} = \alpha_i \bar{q}_s$ is applied to the nonlinear continuous-time system (4)-(5) and controlled with sampling time $\tau$. Figure 8 shows that constraints are correctly enforced for each subsystem.

## 5. CONCLUSIONS

In this paper a hierarchical control scheme has been proposed for the coordination of an ensemble of steam generators, which must cooperate to fulfill a common load. The definition of an ensemble reference model, as proposed here, permits to solve the medium level tracking MPC in a scalable and flexible way, as its dimension does not grow with the number of steam generators in the ensemble. Thanks to the model reformulation, the ensemble model can be simply obtained from the solution to the high level problem and updated online. The model configuration is determined by the high-level bilinear mixed-integer optimization that computes the optimal number of generators to be included in the ensemble and their shares of steam production by minimizing the operating cost and considering global and subsystem constraints. Moreover, at this level, the demand is considered as an optimization variable to avoid feasibility problems at medium level. Future work will consider the improvement of the multi-layer scheme by comparing the overall performance with the implementation of an additional low-level shrinking MPC control to further address the local model mismatch. We also envision to extend the high level optimization including the ensemble dynamics.